\documentstyle[prl,aps,psfig,multicol]{revtex}
\begin{document}
\draft
\title{$H/T$ Scaling of the Magnetoconductance near the 
Conductor-Insulator Transition\\ in Two Dimensions.}
\author{D.~Simonian, S.~V.~Kravchenko, and M.~P.~Sarachik}
\address{City College of the City University of New York, New York, New York 
10031}
\author{V.~M.~Pudalov}
\address{Institute for High Pressure Physics, Troitsk, 142092 Moscow District, 
Russia}
\date{\today}
\maketitle
\begin{abstract}
For an electron density near the $H=0$ insulator-to-conductor transition,
the magnetoconductivity of the low-temperature conducting phase in 
high-mobility silicon MOSFETs is consistent with the form 
$\Delta\sigma(H_{||},T)\equiv\sigma(H_{||},T)-\sigma(0,T) =
f(H_{||}/T)$ for magnetic fields $H_{||}$ applied parallel to the plane of the 
electron system.  This sets a valuable constraint on theory and provides 
further evidence that the electron spin is central to the anomalous $H=0$ 
conducting phase in two dimensions.
\end{abstract}
%\makeatletter
%\global\@specialpagefalse
%\def\@oddhead{REV\TeX{} 3.0\hfill Final version}
%\let\@evenhead\@oddhead
%\makeatother
%\narrowtext
\begin{multicols}{2}

Recent experiments\cite{dsimon,pudal} have demonstrated that the anomalous 
conducting phase\cite{krav} found in the absence of a magnetic field in 
two-dimensional electron systems in silicon 
metal-oxide-semiconductor field-effect transistors (MOSFETs) 
is strongly suppressed by an in-plane magnetic field, 
$H_{||}$.  For electron densities near the $H=0$ transition ({\it i.e.} for 
$\delta n \equiv (n_s-n_c)/n_c<<1$, $n_c \sim 
10^{11} \mbox{cm}^{-2})$, an external 
parallel field as low as $H_{||}\sim$4 kOe 
causes an increase in the resistivity, changing the 
sign of $d\rho(T)/dT$ at low temperatures from positive (metallic) to negative 
(insulating behavior); the resistivity saturates to a constant value for 
fields $H_{||}$ above $\sim$20~kOe, indicating that the conducting phase 
has been entirely quenched.  We have shown further\cite{nature} that 
a magnetic field suppresses the conducting phase 
independently of the angle of application with respect to the 
2D electron layer.  
The total magnetoconductance is the superposition of this term and 
orbital effects associated with the perpendicular component of 
the field which give quantum Hall oscillations\cite{highns}.   

The unexpected conducting phase in two dimensions in the absence of a 
magnetic field has been observed recently for holes in SiGe quantum 
wells\cite{coleridge} and GaAs/AlGaAs 
heterostructures \cite{exeter,yael,pepper}.  
Although considerably smaller, a negative magnetoconductance (positive
magnetoresistance)\cite{pepper,private} found in these systems has 
also been attributed to the suppression of the 
conducting phase.

An in-plane magnetic field affects the spins of the electrons only, and has 
no effect on their orbital motion.  The quenching of the conducting phase by 
a magnetic field applied parallel to the plane of the electrons thus provides 
strong indication that the electrons' spins play a central role in the 
anomalous conducting phase in these two-dimensional systems.  We now 
demonstrate that near the metal-insulator transition, the magnetoconductivity 
of the $H=0$ conducting phase in high-mobility dilute silicon MOSFETs scales 
with $H/T$, obeying the form
\begin{equation}
\Delta\sigma(H_{||},T)\equiv \sigma(H_{||},T)-\sigma(0,T) = 
f(H_{||}/T).
\label{eq}
\end{equation}
\vbox{
%\vspace{.05in}
\hbox{
\hspace{.4in}
\psfig{file=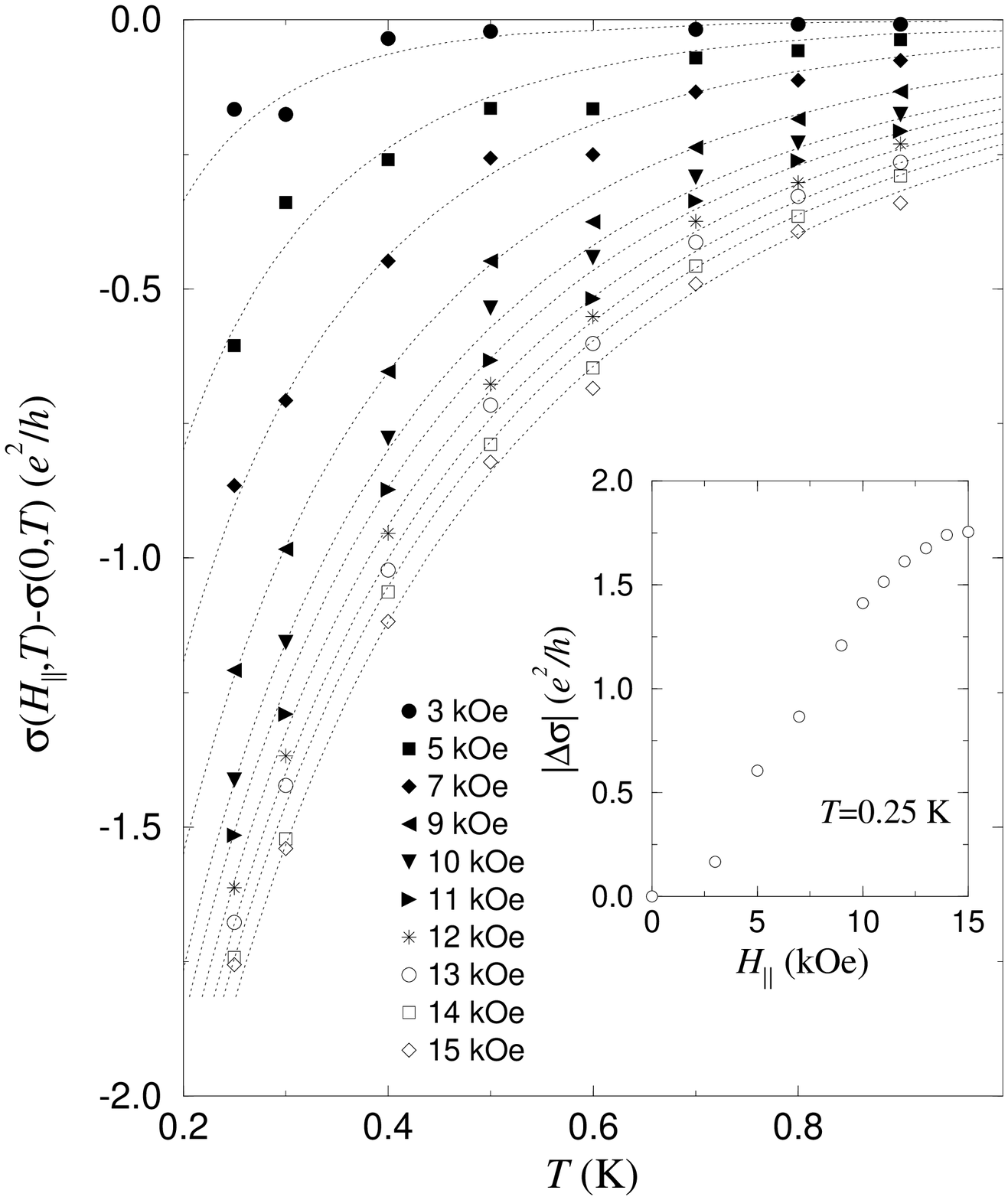,width=3.1in,bbllx=1.5in,bblly=1in,bburx=7.75in,bbury=9.25in,angle=0}
}
\vspace{0.15in}
\hbox{
\hspace{-0.15in}
\refstepcounter{figure}
\parbox[b]{3.4in}{\baselineskip=12pt \egtrm FIG.~\thefigure.
Magnetoconductivity $\Delta\sigma(H_{||},T)\equiv\sigma(H_{||},T)-
\sigma(0,T)$ versus temperature $T$ for several magnetic fields $H_{||}$ 
applied parallel to the plane of the electrons. The sample is in the 
conducting phase with an electron density $\delta = (n_s-n_c)/n_c=0.10$. 
The dotted lines are guides to the eye.  The inset shows the absolute value 
of the magnetoconductivity $|\Delta\sigma(H_{||},T)|$ versus $H_{||}$ 
at the lowest measured temperature, $T=$0.25 K. Note the rapid increase of 
$|\Delta\sigma(H_{||},T)|$ followed by saturation above $\approx13$~kOe.
}
\label{fig1}
}
}

The maximum mobility of the sample used in these experiments 
was $\mu^{max}_{T=4.2K}\approx25,000$~cm$^2$/Vs.  The conductivity was 
measured in magnetic fields up to 15 kOe applied parallel to the plane of the 
electrons. 
\vbox{
\vspace{0.4in}
\hbox{
\hspace{0.4in}
\psfig{file=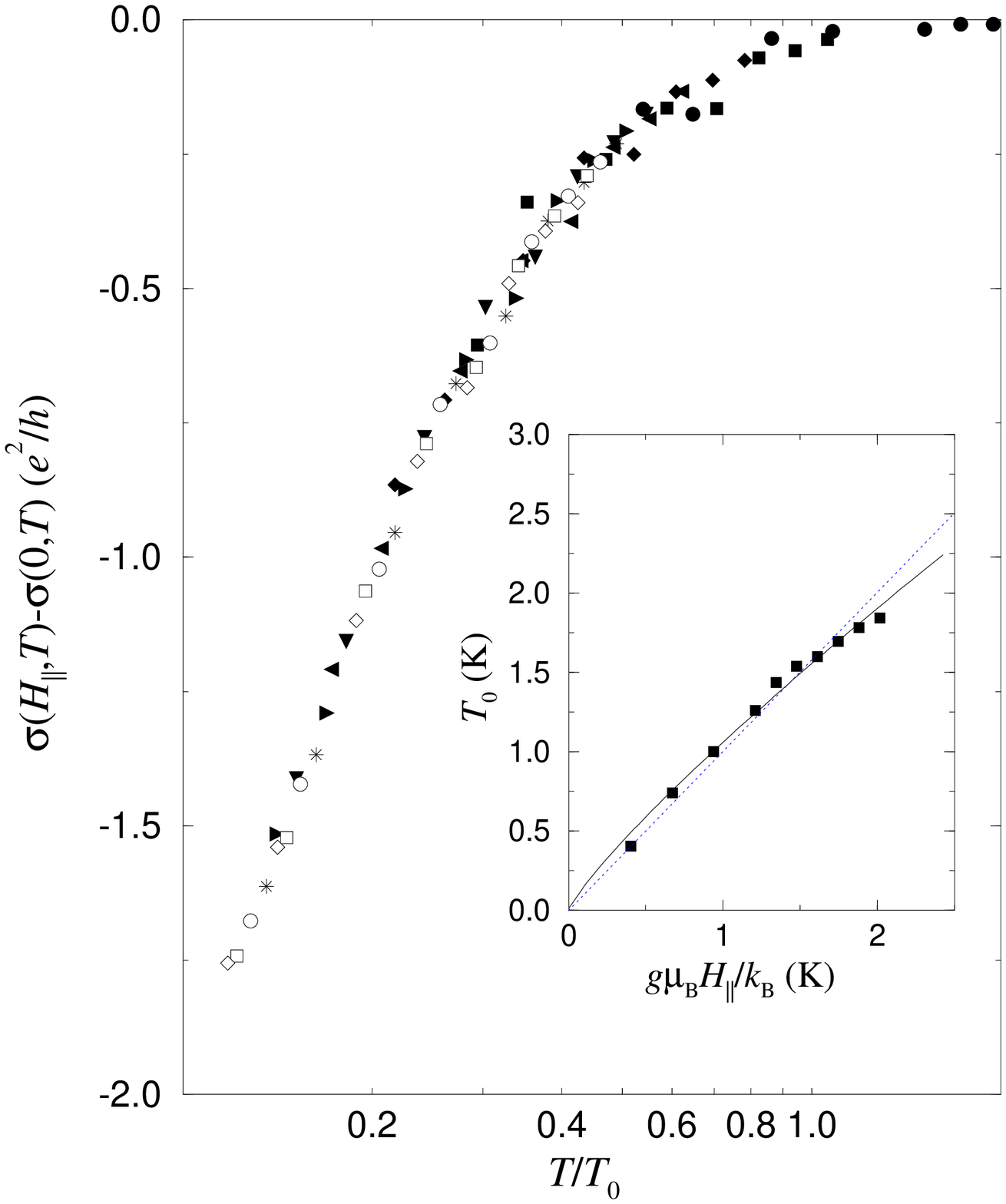,width=3.1in,bbllx=1.5in,bblly=1in,bburx=7.75in,bbury=9.25in,angle=0}
}
\vspace{0.15in}
\hbox{
\hspace{-0.15in}
\refstepcounter{figure}
\parbox[b]{3.4in}{\baselineskip=12pt \egtrm FIG.~\thefigure.
The magnetoconductance $\Delta\sigma$ as a function of $T/T_0$.  
The inset shows the scaling parameter $T_0$
plotted as a function of $g\mu_BH_{||}/k_B$. (Symbols for
different fields, $H_{||}$, are the same as in Fig.~\ref{fig1}). 
A power-law fit, 
shown by the solid curve, 
yields $T_0 \propto H_{||}^\alpha$ with $\alpha={0.88\pm 0.03}$.  The dotted
straight line corresponds
to $T_0=g\mu_BH_{||}/k_B$; deviations from straight-line behavior are 
attributed to saturation at high fields.
\vspace{0.10in}
}
\label{fig2}
}
}
Measurements were taken between $0.25$ and $0.9$ K with the 
sample immersed in the $^3$He-$^4$He mixing chamber of a
dilution refrigerator.  
The electron density was set by the gate voltage at
$n_s=9.43\times10^{10}\text{cm}^{-2}$, placing the 
sample on the conducting side and near the conductor-to-insulator 
transition ($n_c=8.57\times10^{10}\text{cm}^{-2}$). In the absence 
of a magnetic field, the resistance was 
$13.9$ kOhm at the lowest measured temperature, $T=0.25$~K.

The magnetoconductivity, $\Delta\sigma(H_{||},T)=\sigma(H_{||},T)-
\sigma(0,T)$, is shown in Fig.~\ref{fig1} as 
a function of temperature for various fixed values 
of parallel magnetic field.  
The magnetoconductance is negative, its absolute value 
increasing with applied field and 
with decreasing temperature.  The noise for small $H_{||}$ derives from the 
subtraction of two large (and comparable) quantities, 
$\sigma(H_{||},T)$ and $\sigma(0,T)$.  The inset to Fig.~\ref{fig1} shows the 
absolute value of the magnetoconductivity as a function of $H_{||}$ at a 
temperature of $0.25$K. The absolute value of magnetoconductance rises rapidly 
and begins to saturate above $\sim13$~kOe, consistent with earlier 
measurements\cite{dsimon,pudal}.  The data for
\vbox{
\vspace{1mm}
\hbox{
\hspace{0.10in}
\psfig{file=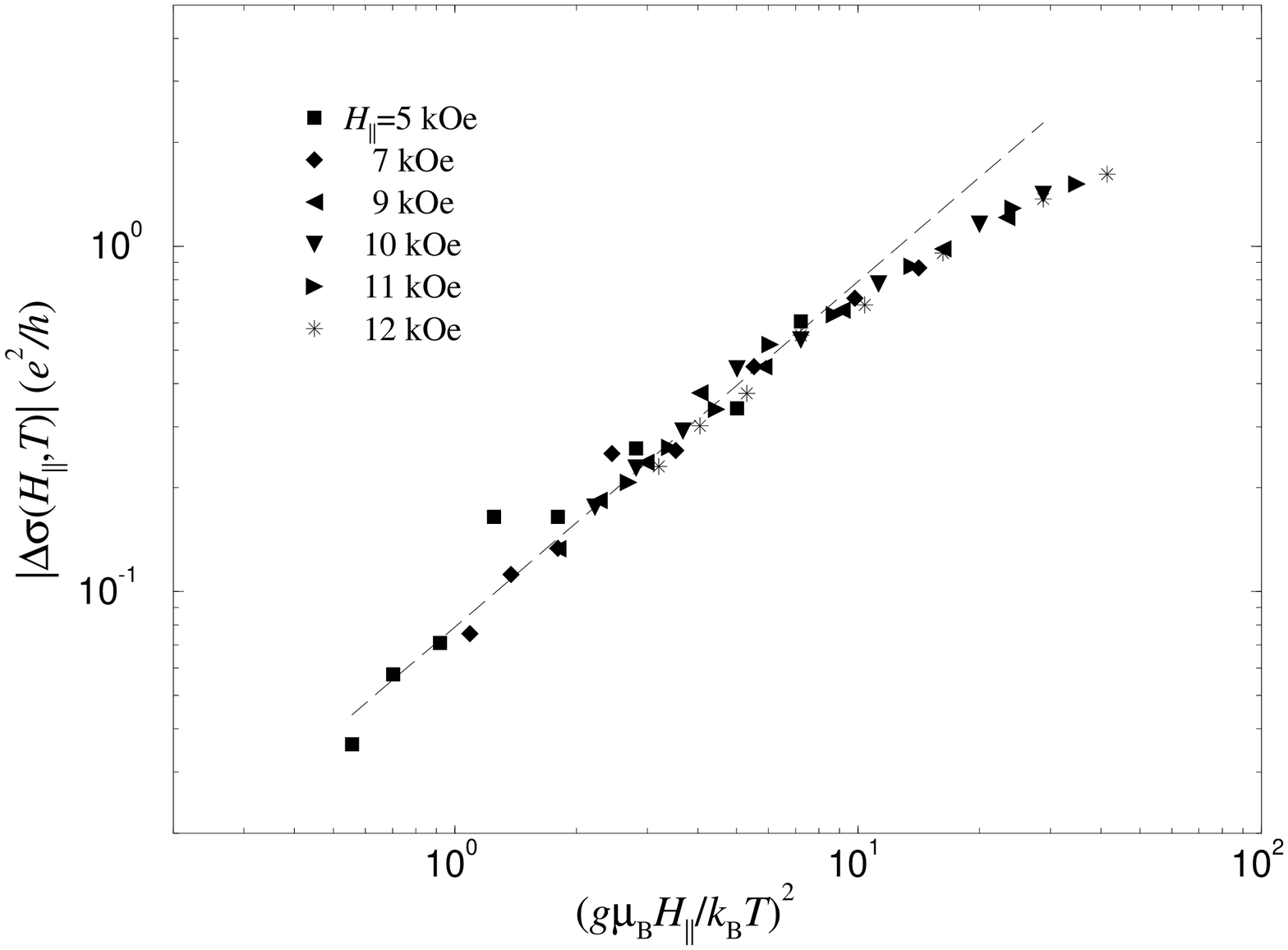,width=3.3in,bbllx=.5in,bblly=1.25in,bburx=7.25in,bbury=9.5in,angle=-90}
}
\vspace{0.3in}
\hbox{
\hspace{-0.15in}
\refstepcounter{figure}
\parbox[b]{3.4in}{\baselineskip=12pt \egtrm FIG.~\thefigure.
Magnetoconductance $\Delta\sigma(H_{||},T)$ versus $g\mu_BH_{||}/k_BT$ 
on a logarithmic scale for $H_{||}=5~\text{to}~12$ kOe (see text).  The dashed
line is a fit to Eq.~7 in Ref.\cite{castellani}, 
$\Delta\sigma(H,T)=-0.084e^2/(\pi h)\gamma_2(\gamma_2 +1)(g\mu_BH/k_BT)^2$. 
$\gamma_2=1.3$.\vspace{0.10in}
}
\label{fig3}
}
}
$\Delta\sigma$ can 
be collapsed onto a single curve by applying a different multiplicative factor 
to the abscissa for each curve, as illustrated in Fig.~\ref{fig2}.  The 
scaling parameter $T_0$ is plotted in the inset as a function of the Zeeman 
energy, $g\mu_BH_{||}/k_B$ (in Kelvin).  Here  $g$ is the $g$-factor (equal to 
2 in Si MOSFETs), $\mu_B$ is the Bohr magneton, and $k_B$ is the Boltzmann 
constant.  A power-law fit (shown by the solid curve) yields $T_0\propto 
H_{||}^\alpha$, with $\alpha=0.88\pm0.03$.  We note that $H/T$ scaling of the 
form Eq.~(\ref{eq}) requires that $\alpha=1$, corresponding to $T_0= 
g\mu_BH_{||}/k_B$ (indicated in the inset by the dotted line).  
We suggest that the deviation of $\alpha$ from unity is associated with 
the saturation of the magnetoconductance at $H_{||}\gtrsim13$~kOe shown in the 
inset to Fig.~\ref{fig1}, where one might well expect the scaling to break 
down.  We therefore exclude the data sets at the three largest fields (for 
which the proximity of the scaling parameter $T_0$ to saturation is apparent). 
For in-plane fields in the range $H_{||}$=5...12 kOe, the absolute value of 
magnetoconductance, $|\Delta\sigma(H_{||})|$, is shown as a function of 
$(g\mu_BH_{||}/k_BT)^2$ in Fig.~\ref{fig3}.  For this range of magnetic fields 
and for an electron density fairly close to the critical density, the 
magnetoconductance scales well with $H_{||}/T$.

Based on quite general arguments, Sachdev\cite{sachdev} showed that the 
conductivity near a second-order quantum phase transition is a universal 
function of $H/T$ for a system with conserved total spin.  If the 
magnetoconductance of our silicon MOSFET does indeed scale with $H/T$ 
(the dotted straight line in the inset to Fig.~\ref{fig2}) rather than 
$H/T^\alpha$ with $\alpha\neq1$ (the solid curve), this would imply that 
spin-orbit effects 
are relatively unimportant near the transition.  For a weakly interacting 
2D system, Lee and Ramakrishnan\cite{rev85,leerama} obtained scaling of the 
form Eq.(\ref{eq}) associated with the negative $|S_z|=1$ triplet channel 
contribution to the conductance.  We note that the scaling reported here 
for the 2D system in silicon MOSFETs is remarkably similar \cite{sim} 
to the $H/T$ scaling of the magnetoconductance observed by Bogdanovich 
{\it et al.}\cite{snezana} in three-dimensional Si:B near the metal-insulator 
transition, where it was attributed to the mechanism of Ref.\cite{leerama}.  
$H/T$ scaling is also expected within 
theories that predict various types of superconductivity in a strongly 
interacting system in two dimensions\cite{belki,phili,rice}.

Extending earlier work of Finkel'shtein\cite{sasha}, who showed that a 
disordered, weakly interacting 2D system can scale toward a metallic phase, 
Castellani {\it et al.}\cite{castellani} have obtained a 
magnetoconductance $\Delta\sigma(H,T)=-0.084e^2/(\pi h)
\gamma_2(\gamma_2 +1)(g\mu_BH/k_BT)^2$. The coupling constant $\gamma_2$ is 
expected to vary with temperature in the range of validity of the calculation, 
namely, not too close to the critical density.  
Our observation of simple $H/T$ scaling for a relative density $\delta_n\ll 1$ 
implies that $\gamma_2$ is at most a weakly temperature-dependent quantity 
near the transition. The fit to the form suggested in Ref.\cite{castellani}
is shown by the dashed line in Fig.~\ref{fig3}, and yields $\gamma_2 
\approx1.3$ ($\gamma_2(\gamma_2+1)\approx3$), corresponding to intermediate 
coupling strength~\cite{palpriv}.

To summarize, we have observed scaling of the magnetoconductivity of the 
form $\Delta\sigma(H_{||},T)=f(H_{||}/T)$ in the anomalous conducting phase 
of a two-dimensional system of electrons near the conductor-to-insulator 
transition.  In this, as in other systems where a conducting phase has been 
observed at low temperatures in the absence of a field, estimates
\cite{krav,coleridge,exeter,yael,pepper} indicate that the energy of 
interactions 
between carriers is much larger than the Fermi energy in the range of carrier 
densities where the conducting state exists.  The suppression of the 
conductivity by an in-plane magnetic field is consistent with a decrease 
of the spin-dependent part of the electron-electron interactions.  

Numerous suggestions have been made regarding the nature of the unexpected 
conducting phase in two dimension in the absence of a field, and a consensus 
has yet to emerge.  Our data provide further evidence that the spins play 
a central role.  Moreover, our finding that the magnetoconductance in the 
conducting phase near the transition scales with $H/T$ sets a 
valuable constraint on theory.

We are grateful to Lenny Tevlin and Subir Sachdev for independently suggesting 
the scaling analysis presented in this work.  We thank D.\ Belitz, J.\ L.\ 
Birman, Song He, D.\ I.\ Khomskii, P.\ A.\ Lee, P.\ Phillips, 
T.\ M.\ Rice, T.\ V.\ Ramakrishnan and F.\ C.\ Zhang for valuable discussions. 
This work was supported by the US Department of Energy under 
Grant No.~DE-FG02-84ER45153. V.\ P.\ was supported by RFBR(97-02-17387) and by 
INTAS.

\end{multicols}

\begin{references}
\bibitem{highns} The response to a perpendicular field is considerably more
complicated and has been studied by the authors of Refs.\cite{pudal}, 
\cite{nature}, and \cite{dragana}.
\bibitem{sim} Similarities between the two systems were pointed out in 
Ref.\cite{dragana}.
\end{references}
\end{document}